# Lagrangian Coherent Structures (LCS) may describe evolvable frontiers in natural populations

**Bradly Alicea**
**Department of Animal Science, Michigan State University**
**bradly.alicea@ieee.org**

## Abstract

The evolution of organismal populations is not typically thought of in terms of classical mechanics. However, many of the conceptual models used to approximate evolutionary trajectories have implicit parallels to dynamic physical systems. The parallels between currently-used evolutionary models and a type of model related to Lagrangian Coherent Structures (LCS) will be explored. The limits of evolvability in a population can be treated in a way analogous to fronts, waves, and other aggregate formations observed in fluid dynamics. Various measures and architectural features will be introduced. Relevant scenarios include so-called evolvable boundaries and related scenarios involving evolutionary neutrality, such as migrations, demographic bottlenecks, and island biogeography. The LCS-like model introduced here could eventually be applied to a wide range of problems that normally utilize forms of evolutionary modeling.

## Introduction

Evolutionary processes and their major features have commonalities with physical dynamical systems. In terms of inspiration and major assumptions, Models of fitness and adaptation currently used in theoretical biology are in many ways unintentionally similar to physical dynamical models known as a Hamiltonian. However, the Hamiltonian metaphor may not be appropriate for every evolutionary scenario. In this paper, the limits of evolvability and related neutral processes for a given population will be treated as an emergent phenomenon that can conceptualized in a manner similar to fronts, waves, and other aggregate formations observed in collective animal behavior [1] and fluid dynamics alike. Thus, the case will be made that evolving populations can also be conceptualized using a computational approach based on a mathematical tool called Lagrangian Coherent Structures (LCS).

To place this observation in context, a quick overview of how Hamiltonian-like models (fitness landscapes and hypercubes) are currently used to approximate the relationships between genotypes/phenotypes, fitness, and evolvability among populations will be presented. An underlying theme involves the implicit relationship between evolutionary models and Hamiltonian dynamics. As an alternative, an LCS-like model with similarities to agent-based approaches will be proposed. Essential features of this model, relevant evolutionary scenarios, and comparisons to Hamiltonian-like models will be discussed, along with a quick summary of future research directions.

## Evolutionary Modeling as a Continuous Dynamical Phenomenon

One premise is that physical models are appropriate for representing evolutionary systems given they both consist of similar features. A concurrent theme in current evolutionary theory is that evolution is an algorithmic process [2]. Thus, maximization of parameters related to adaptation (e.g. fitness) can be understood using continuous, *n*-dimensional space called a landscape. Biologists in the early 20th century such as Wright [3] and Waddington [4] used such landscapes as a qualitative tool. Fitness landscape theories have been advanced by people like Kauffman (*nk*-landscapes, [5]), Gavrilets (holey adaptive landscapes, [6]), and Wagner (hypercube networks, [6]). Hamiltonian systems share two attributes with these types of evolutionary models: accessibility with respect to initial condition, and optimization of outcomes over time [8, 9].

In the case of a first-order Hamiltonian, particle populations evolve from an ergodic distribution, or space where every portion of the space is equally probable, to a heterogeneous spatial configuration characterized by the minimization of energy [10]. Ergodicity can be contrasted with representations of variation in fitness, where some configurations are clearly more accessible than others. Protein-folding simulations provide an intermediate example of fitness landscapes and physical Hamiltonians, as the most energetically efficient protein topologies also tend to possess the highest fitness. In molecular dynamics simulations, kinetic profiles resulting from random movement of particles reveal an optimal point in the distribution where the greatest amount of free energy is conserved [11]. This profile can be extended to multiple dimensions, each of which has an optimum, much like current representations of fitness. However, the solution relies upon the discovery of optimal points, which are sometimes hard to access.

The second premise suggests that by understanding the connections between evolutionary constructs and dynamical physical systems, we can arrive at more specific evolutionary approximations. Such topological models focus with the relationship between genotype and phenotype, such as the characterization of many potential routes to fitness maxima [5]. Placing constraints on pathway number and diversity over time is how the limits of evolvability, or the collective capacity of individuals in a population to respond to environmental challenges [12], is characterized in these evolutionary representations. While fitness landscapes are organized around the idea of optima, fitness functions themselves are shaped by changes in the fitness values of a population. Thus, current evolutionary representations tend to emphasize optimal, mean field properties rather than boundary conditions and other less prominent features of evolutionary dynamics [13, 14]. Given these constraints, a phenomenon such as the limits of evolvability may be hard to interpret using a Hamiltonian-based approach.

## Lagrangian systems: an alternative model for evolutionary systems

In what ways can be improve upon current evolutionary representations, while still retaining the formal mathematical underpinnings of physical models? Lagrangian systems, while also based on classical mechanics, differ from Hamiltonian systems in two ways. The first involves evaluating the kinematic profile of a dynamic process while holding its kinetic properties constant. Representing evolutionary systems in this way does not require energy minimization nor fitness maximization, and allows for particles representing individuals to be monitored in a topological space.

Secondly, the path of a particle in Lagrangian systems is inferred from observations of the initial condition, observations of discrete measurements, and a constant energy function that represents diffusion. While they cannot be directly predicted at any single point in time, meta-observations such as an evolutionary trajectory or evolutionary constraints can be inferred. The formation of Lagrangian Coherent Structures in this type of mean field model allows us to place evolutionary process in the context of a geographic analogue, which may be crucial to approximating a number of population dynamics scenarios.

**Basic Topology for Lagrangian Evolution**

As a computational abstraction for evolutionary dynamics, the LCS-like approach has many unique features. One unique feature is a multidimensional volume through which entities traverse during the process of evolution. Instead of being concerned with the path an organism takes and whether it leads to higher fitness, the focus is now on the patterns that results from a fluid dynamics like process which serves as a stand-in for environmental selection and constraints. An emerging, second-order formulation of Lagrangian dynamics called Lagrangian Coherent Structures (LCS) might provide a suitable framework for investigating these properties. To model an evolutionary system using a second-order Lagrangian approximation, there are two basic components that must be defined. The first of these is rate of change, which relates to how fast entities are evolving. The second of these is direction of change, which corresponds to the direction in which things are evolving and how many members of a population are headed in that direction. These parameters will also be used to map collective movement to evolutionary dynamics. As these measures are suitable for uncovering dynamics related to phenomena such as evolutionary neutrality and evolvablity, using a Lagrangian method as the basis for an evolutionary model becomes more plausible.

**Lagrangian Coherent Structures for Evolutionary Representation**

LCS-like evolutionary models describe a different set of phenomena than do models that present evolution as an exercise in fitness maximization. In cases where the Hamiltonian-like landscape is exceedingly rugged [5], it can be hard to represent evolution of adaptive traits even for small mutational changes. In the course of hill-climbing, properties of evolutionary systems such as evolvability become epiphenomenal. While a similar type of objective occurs in the LCS-like model, movement across the space is more closely tied to differential survival and diffusive processes. To demonstrate this, more detail of the architecture that parameterizes evolvability instead of approximating a route to maximum fitness is required.

Lagrangian Coherent Structures [15] have been applied to mechanical [16] and ecological [17] problems in biology that explicitly involve movement and the interactions between objects and the flow fields that surround them. It has been proposed [18] that an LCS-like model (in this case, a related hybrid model) can be used to represent the structural features that underlie the evolutionary dynamics of culture. With modifications to the model and representation, the same approach will be used here. Thus, a rough sketch of a representation for a range of evolutionary dynamics will now be presented.

## Structure/Function of the LCS-like Evolutionary Model

As we will see, the basic LCS-like model can be configured in many ways. In this sense, it is extensible to a host of evolutionary system contexts. To demonstrate how this model works, I will review a multi-stage cartoon (schematic diagram) shown in Figure 1 (frames A-E). A process diagram of how evolvable frontiers emerge is demonstrated in Figure 2. In frame A, a genotype-phenotype relationship is mapped to a 3-dimensional Euclidean space prior to the initial condition. Frame B shows that a population of particles all originating at the same position in this space is used to initialize the model.

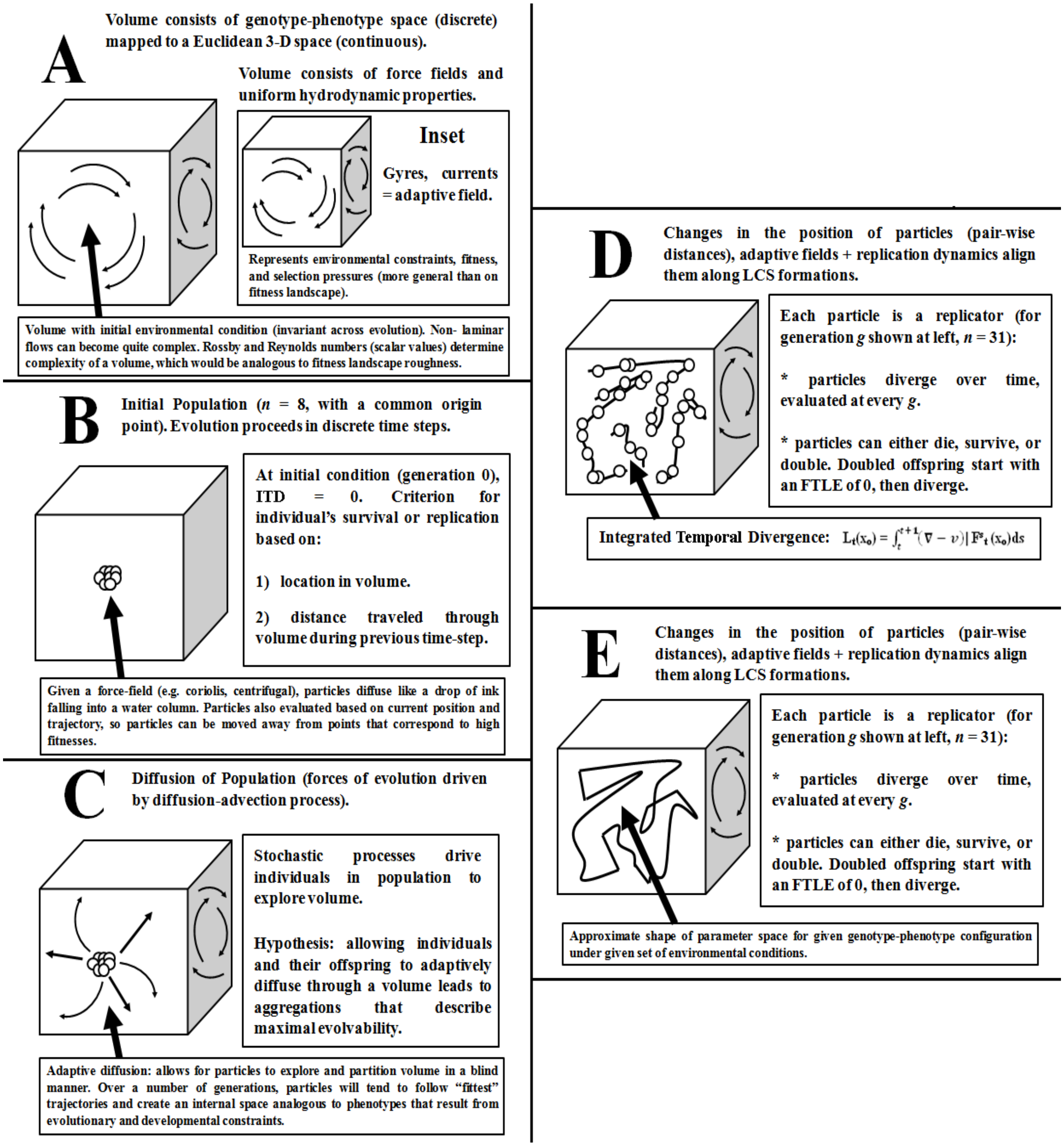

**Figure 1. Cartoon depicting evolvable volume architecture over time.**

Each particle contains a genomic representation, which can be either homogeneous or heterogeneous across the population (see Supplemental Materials - Methods - Genomic Representation). This particle population then diffuses according to a diffusion-advection process (see Figure 1, frame C). To create variable conditions for dispersal, the flow field is governed by a series of deterministic flow jets (see Supplemental Materials - Methods - Flow Field) that can be configured to create potentially turbulent conditions.

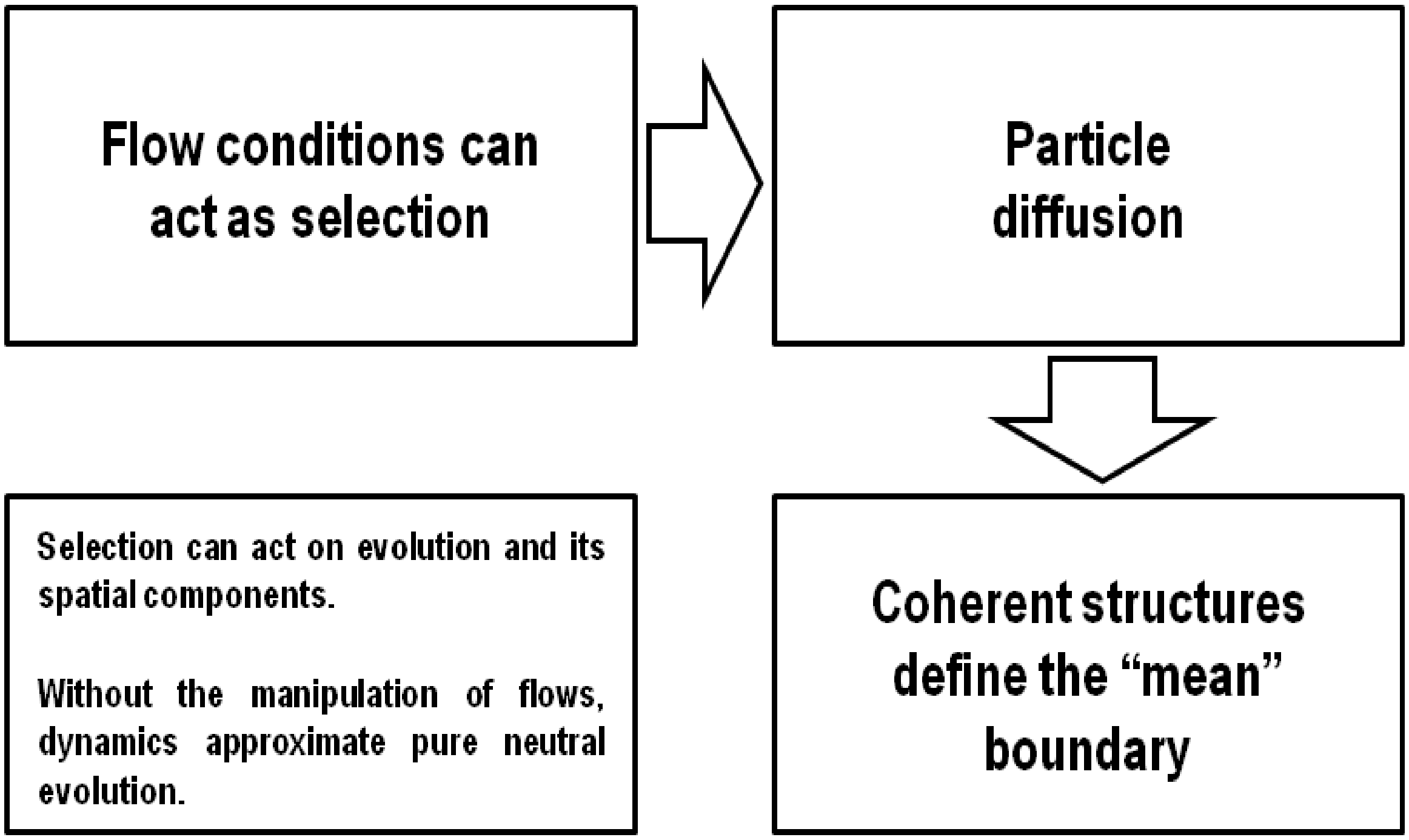

**Figure 2. Outcomes and Conditions for the LCS Model of Evolutionary Dynamics.**

Once particles have diffused across the volume, the iterated temporal divergence measure (ITD - see Supplemental Materials - Methods - Measures) can be used to evaluate particles at distinct points in time or space. Based on the scenario in Figure 1, frame D, we can see that the Finite-time Lyapunov Exponent (FTLE – congruent to the ITD measure) is zero at the initial condition and a value that characterizes evolutionary divergence at the time of measurement. In frame E, we can see that particles have the potential to differentially replicate, survive, and ultimately cluster together given their encountered flow conditions and internal genomic representation. The ITD measure can be calculated pairwise for all particles in the population in parallel, which can yield these higher-level patterns across the entire volume.

Models based on Lagrangian dynamics also allow us to examine some of the more subtle features of evolutionary systems. Using the same configuration as in Figure 1, additional measures such as the segregation factor (see Supplemental Materials - Methods - Measures) and conditional diversity (see Supplemental Materials - Methods - Measures) can be used to assess the diversity (e.g. genomic content) of particles at nearby positions (e.g. same clusters or ridges) at time $t$.

Supplemental Figure 1 is a schematic diagram that demonstrates the various expected coherent structures and diverse outcomes of the model. Yet the LCS-like model can also provide an informative approximation specific to an organismal (individual) or population (collective) context. In Figure 1, much like in a fitness landscape, the genotype-phenotype relationship is shown to be fairly vague and abstract. While evolutionary features can be represented using an LCS-like model, these models can also be configured for providing useful information about functional and structural relationships.

Returning to the general architecture shown in Figure 1, let us recall that each particle exhibits evolutionary dynamics, replicator dynamics, and the ability to interact with fluid dynamics. Specifically, each particle is a replicator (see Supplemental Materials - Methods - Replicator Dynamics) that can die, survive, or double based on the survival measurement taken at specific intervals (see Figure 3). When particles double, the offspring diverge from the parent from the generation of replication. These replicator dynamics are contingent upon environmental (e.g. flow) conditions, but could also be a pre-programmed response of the genome. In terms of evolutionary time, diffusive processes drive particles throughout the volume. These diffusive processes resemble neutral processes in evolution [19], which may result in neutral processes dominating the flow field representation. However, the introduction of alternate flow conditions may also act exclusively as a selective force. Given the current design of LCS-like models, there is no way to separate out bona-fide selective forces from other evolutionary dynamics. However, this is not beyond the scope of LCS-like models.

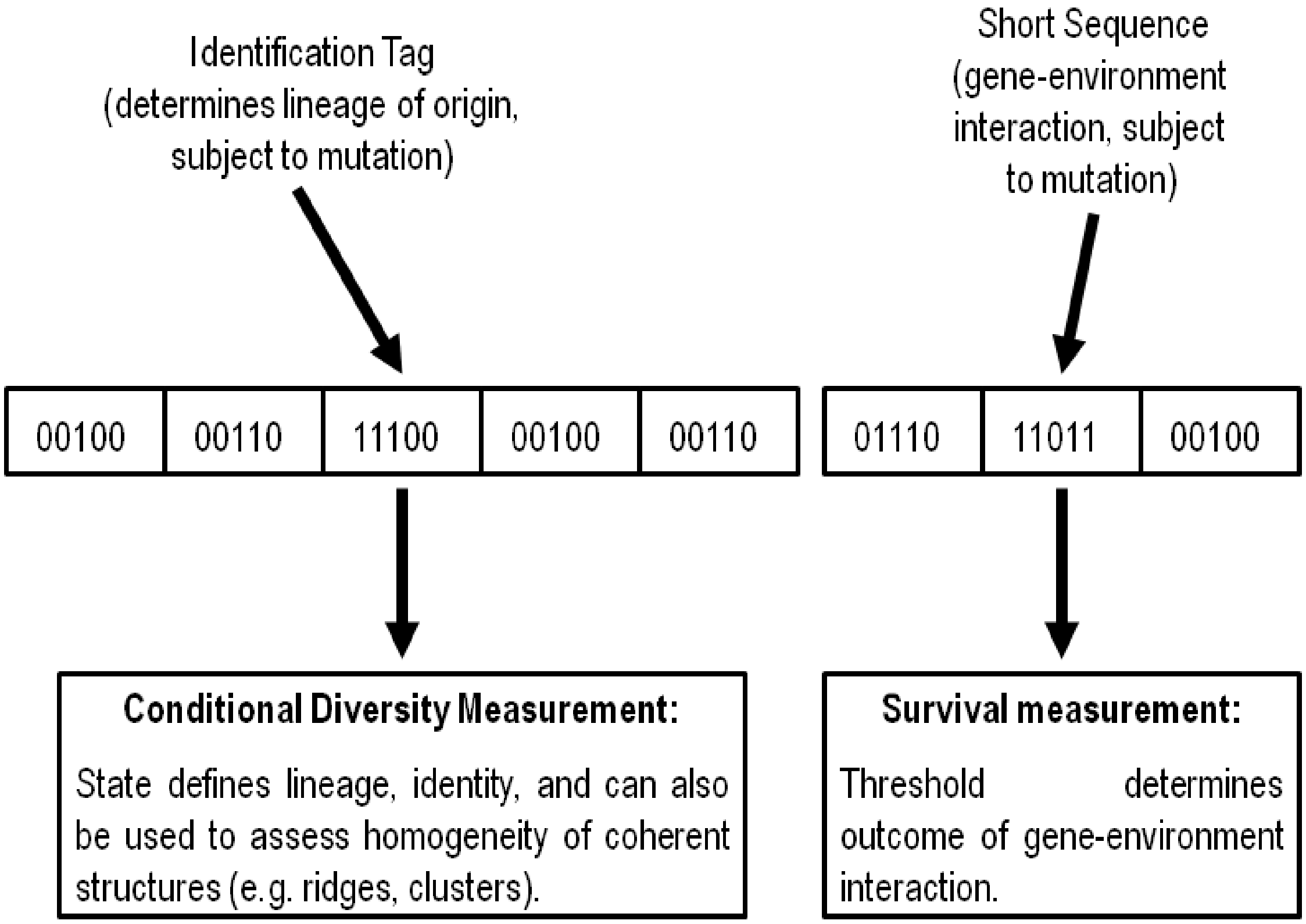

**Figure 3. Architecture of a single particle genome.**

**Approximating Rates and Directions of Evolutionary Change**

One set of features that can be well-characterized using the LCS-like model are the *rates* and *directions* of evolutionary change. For theoretical populations, this parameter can be explicitly defined and compared across contexts using two variations on the ITD measure: the finite-time Lyapunov exponent (FTLE, which approximated the rate of evolutionary change), and the finite-space Lyapunov exponent (FSLE, which approximated the direction of evolutionary change).

**Evolutionary Rate of Change.** To characterize the rate of change in a hypothetical evolutionary system, we can use an existing variable from the LCS literature called the finite-time Lyapunov exponent (FTLE). This parameter allows us to measure the distance a particle lineage (see Supplemental Materials - Methods - Replicator Dynamics) has traveled from its point of origin to a final point of measurement. Recall that this space uses a coordinate system rather than fitness values, which allows us to characterize evolutionary distance with respect to both individual and collective behavior. While the FTLE measure does not allow us to predict an exact trajectory from the initial condition, it does result in a portrait of how differential responses to environmental variation can lead to similar (or very different) evolutionary outcomes. In conjunction with information from the identity portion of the particle genome, these distances can be used to assess the exploratory potential of both individuals and groups of particles. In this manner, we can approximate evolutionary changes with regard to time.

**Evolutionary Direction of Change.** To characterize the direction of change in a hypothetical evolutionary system, we can use an existing variable from the LCS literature called the finite-space Lyapunov exponent (FSLE). While the FTLE measurement provides a measurement from initial condition to final position at the time of measurement, it does not allow us to predict the actually trajectory a particle takes to get there. While the FSLE measure is similarly unable to provide trajectory information, it can provide distance information about two particle lineages that originate in the same location. This allows us to compare distances between any two particles or aggregates of particles, which may allow us to approximate evolutionary changes that occur with regard to divergence. In this case, divergence can be rooted in a spatial context, which is convenient for purposes of comparisons with respect to changes measured over time. As we will see, approximating the ability to freely diffuse our particles and then measure their diffusion in space is similar to sending out probes to explore all possible configurations for our evolvable system.

**Alternative Configurations of LCS-like Model**

While Figure 1 shows a generalized representation of each dimension, one advantage of the LCS-like model is that each dimension can be explicitly defined. Using a hypothetical example, Figure 4 demonstrates this in the context of four co-expressed genes. In this case, there is an explicit relationship between the independence of each gene and their collective action. There are two interactions here: one between a gene represented on a single dimension and a particle's genomic representation, and another between the volume geometry and the particle's genomic representation. Taken together, these interactions define the physiological action (e.g. co-expression) of all four genes in evolutionary time.

In Figure 5, a more complex example of a LCS-like model with specified dimensions is given. In this case, a six-dimensional volume is shown with both genotypic and phenotypic traits specified on the individual dimensions. Notably, in real biological systems, combinations of traits such as these collectively contribute to the variance seen in a phenotype. However, it is hard to tease out which traits contribute the most to an integrated phenotype, particularly in an evolutionary context. In this formulation, the diffusion patterns of this simulation may provide information about which dimensions explain the most variance.

A demonstration of how boundaries form in a volume is shown in Supplemental Figure 2. These boundaries are based on the ITD measures, and can be thought of as similar to an econometric frontier model [20, 21]. Briefly, the maximal extent of diffusion for all particles forms an evolutionary frontier (see Supplemental Materials - Methods - Frontier Analysis), within which the evolutionary system in question can evolve (e.g. the evolvable space). Multiple instances of the model may reveal different shapes of this frontier, but the general structure is expected to be replicated across instances of the same system.

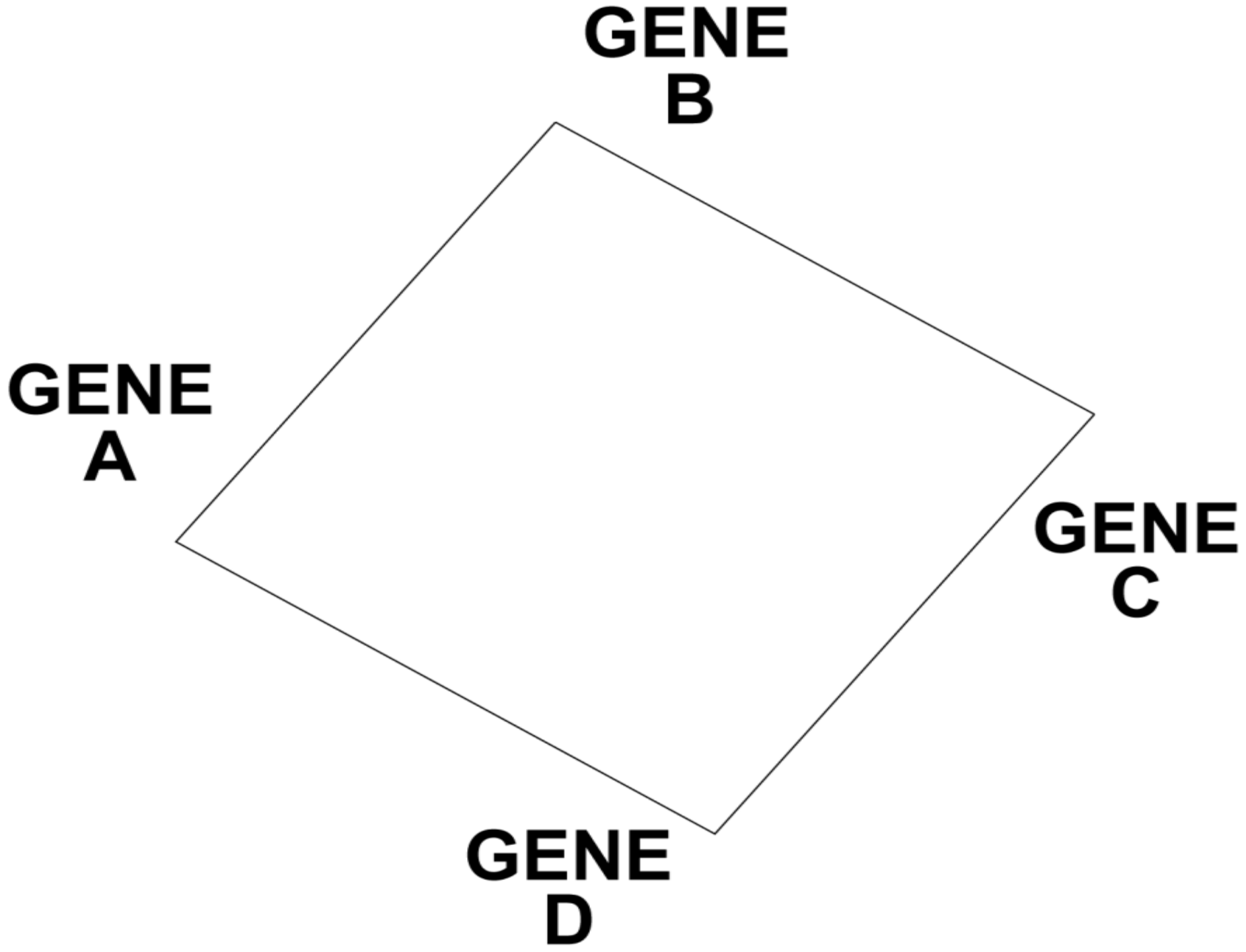

**Figure 4. A four-dimensional flow volume (4-dimensional) that models the synthetic relationship between four co-expressed genes (before particles are seeded).**

**LCS-like Model Applied to General Evolutionary Scenarios**

Our model begins with a population of evolutionary entities that diffuse across a three-dimensional volume at a fixed rate. A given evolutionary trajectory requires a fixed amount of energy, so that selection acts to perturb diffusion. In second-order Lagrangian systems, diffusion of particles in a liquid can be perturbed by a systematic force field. In the LCS-like model, there is no assumption of ergodicity, and environmental conditions imposed by flow fields (see

Supplemental Materials - Methods - Hybrid Fluid Dynamics-Evolutionary Agent Model) can act to actively constrain diffusion via the stochastic dynamics of turbulence. The most likely candidate for this force field in an evolutionary context is an advection-diffusion model (for an example from morphogenesis, see [22]). Once an initial condition is assumed, an evolutionary trajectory will be defined by two factors: the constraints on this trajectory, and trajectory dynamics measured at discrete points in time.

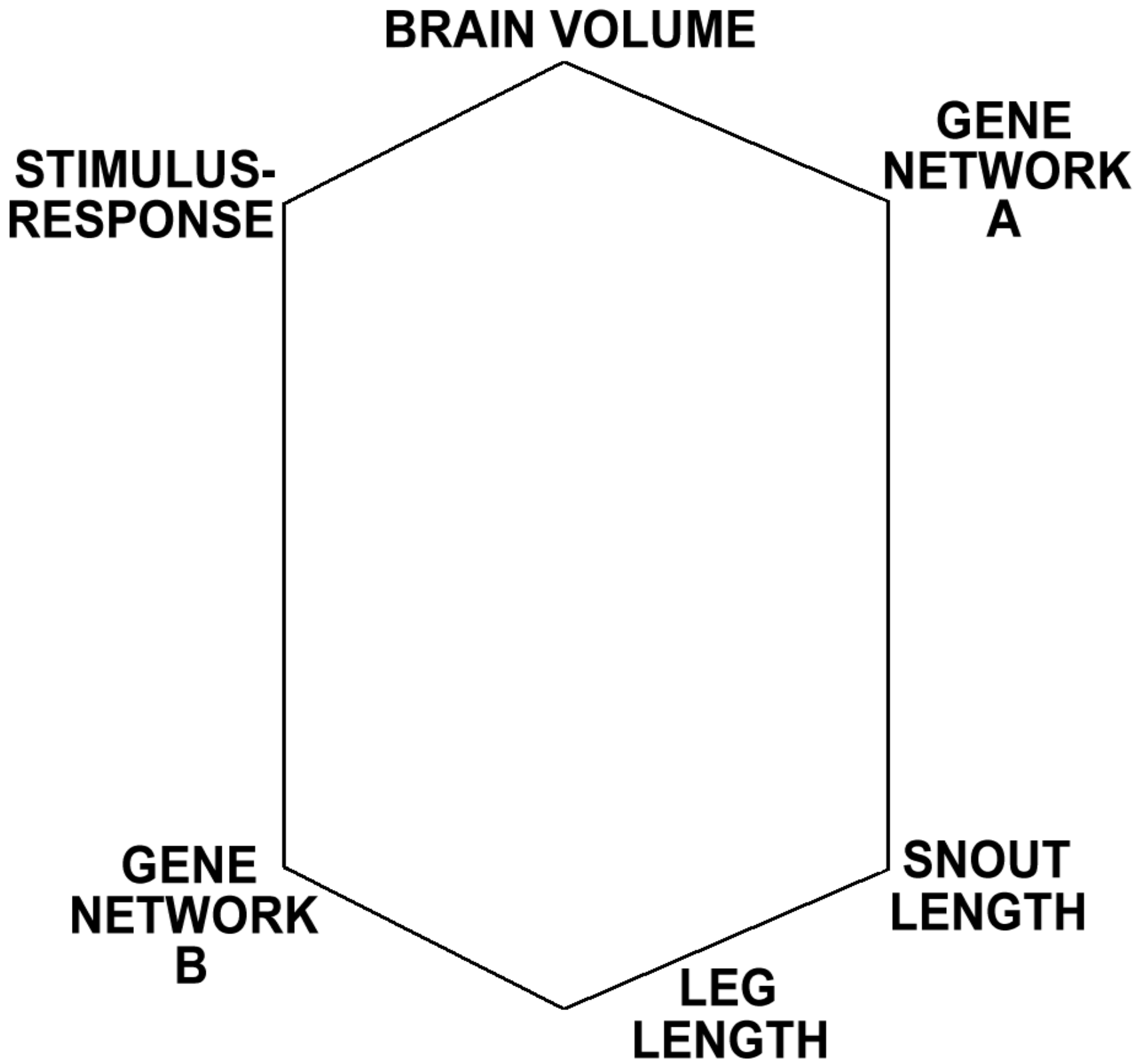

**Figure 5. A six-dimensional volume featuring a variety of specific traits both genotypic and phenotypic (before particles are seeded).**

The constraints on an evolutionary trajectory in a LCS-inspired evolutionary model can also be represented using physical analogues. One specific example is that of forces acting against a subset of potential trajectories, so that certain trajectories are preferred over others. These inertial forces act as dynamic fields, constraining the evolutionary entity as it evolves. In a LCS-inspired model, this imposes a stochastic mechanism for evolutionary constraint.

The evolutionary trajectory of individual particles, each representing individuals in a population, can be measured at discrete points in time using an aggregate parameter. In this context, a formulation of the Lyapunov exponent called the finite-time Lyapunov exponent (FTLE) can be used with respect to evolutionary processes. In general, Lyapunov exponents ($\gamma$) are used in biological dynamical systems to assess the trajectory of a system's state through phase space [23]. In this case of LCS-inspired representations, the FTLE may provide information about the length of a particular evolutionary trajectory.

**LCS-like Models Applied to Evolvability**

Now we turn our attention to the major features of evolvability's limits and how they can be studied in two ways using an LCS-like model. Theoretically, there exists a wide range of ways in which LCS-like models can be applied to studying the limits of evolvability. One way is to use a static approach, which involves modeling interactions between defined morphological structures and an explicitly hydrodynamic environment (see [24, 25]). The other way is to model the diffusion of particles representing an evolving population. In this case, contemporary work on disentangling particle trajectories [26], understanding the effects of geometrical constraints on FTLEs [27], and advanced algorithmic solutions to real-time ridge tracking [28] offer much potential for solving difficult evolutionary problems.

**Examples of LCS-like Model Applied to Evolutionary Neutrality**

Now that it has been demonstrated how evolutionary dynamics play out on this topology, we can review three specific evolutionary scenarios related to neutrality. The relationship between evolvability and neutrality has not been explored extensively at the population level, but at the genotypic level neutral mutations can enable the evolvability of populations [7]. Under some environmental conditions, neutral evolutionary processes can determine the evolvability of a population [29]. It is based on these observations that we now discuss three population-level neutral processes that serve as potential scenarios for future studies: migration, demographic bottlenecks, and island biogeography. The evaluation of rare variants, as a phenomenon related to neutral processes, will also be discussed as an extension of the demographic bottleneck scenario.

**Migration.** Migrations involve the geographic dispersal of populations across a geographic topography. Migration is one process that can be explicitly approximated by flows, as they naturally involve features such as trajectories, velocities, and resistance to boundaries. Over time, we should expect to see coherent structures evolve that are analogous to settlement patterns. By making the initial population heterogeneous, we can introduce constraints on the inherent stochasticity of the migration process. The natural extension of migration dynamics to a LCS-like model allow for two related scenarios (both restricted migrations) to be explored: demographic bottlenecks and island biogeography.

**Demographic Bottleneck.** A demographic bottleneck (e.g. founder effect) is a common feature of populations that are distributed across space (e.g. a varied geography). In this scenario, particles are subsampled in order to re-initialize the volume (see Supplemental Figure 3). After a specified time interval, the new founder population will exhibit a different profile with regard to the original particles. This is particularly the case in terms of their survivability. Physically speaking, the effect is to take an existing cluster and dispersing it in a similar medium. Performing this transfer repeatedly, or returning a previously transferred subsample to its original home, can lead to useful information about the fixation of traits and changes in allele frequencies due to neutral processes.

**Island Biogeography.** Island biogeography is a specific instance of demographic bottleneck, but with an adaptive component. In biological instances of island biogeography, the sequestration of a subpopulation in an isolated location (e.g. an island) leads to highly-specialized adaptations within the subpopulation. In LCS-like instances of island biogeography (see Supplemental Figure 4), the particles are confined to a subspace of the volume via barriers to fluid flow. Physically speaking, the barriers serve to segregate particles and the flows in which they are embodied in. The ability to train subpopulations on partitioned volumes (e.g. islands) effectively serves as both an evolutionary constraint and as a means to uncover the evolvability of subpopulations.

**Rare Variants.** While not tied to neutral processes directly, the persistence or even fixation of rare variants can be driven by neutral processes. Rare variants are sources of genetic diversity that are hard to understand *in vivo*, and have great relevance to the occurrence of disease phenotypes. One way these might be approximated in an LCS-like model is to suspend mutation during evolution and then use a demographic bottleneck scenario to amplify the frequency of selected rare variants in the general particle population. A future goal is to find clusters, ridges, or other structures with genotypes that are rare relative to the rest of the population.

## Conclusions

While approximating evolutionary dynamics using Hamiltonian-style models may be useful, it is an incomplete description of evolutionary systems and processes. LCS-inspired models may complete the picture, particularly in terms of modeling evolvability and the role of neutral processes. Riedl [30] and Schwenk [31] have suggested that evolutionary constraints play a significant role in shaping natural variation. This certainly may be true of neutral processes such as migrations, demographic bottlenecks, and island biogeography.

As conceptualized, the LCS model allows for a population to evolve over time and solutions to aggregate based on the constraints of a given context. As such, evolution over time can result in a dynamically interacting subspace that does not assess fitness *per se*, but does give a direct relationship between the effects of evolutionary processes and the evolvability of the population in the predetermined environment. The boundaries of this process are approximated using a series of measurements, and interpreted using an econometric-style frontier approach. While this can provide significant value as an approximation tool, the predictive value of an LCS-like model applied to evolutionary systems is currently not known.

The application of Lagrangian-like models in evolution, even as a conceptual heuristic, is not without caveats. For example, only those problems that map well to the underlying structure of evolution are likely to perform well using this approach. Problems that are explicitly geographical or involve collective behavior are the best candidates. Nevertheless, by working backwards from physical model to metaphor, we have discovered a method that provides new opportunities for approximating evolutionary dynamics. There is also a lesson to be learned from non-biological natural systems that exhibit evolutionary dynamics (e.g. urban evolution or plate tectonics). In these cases, flows of energy provide a means for structure to be built, which in turn determines the degree of order exhibited by the system [32]. Not only do these types of relationships serve as a template for novel evolutionary representations, but also serve as ready-made scenarios for the LCS-like model.

In conclusion, it must be stressed the message of this paper is not that Hamiltonian-like models be replaced wholesale, but rather that Lagrangian systems might be more informative for a select class of evolutionary problems. With the development of LCS-like models, a new evolutionary analogy has been established. The model may be broadly applicable to open questions in theoretical biology. With modifications, the basic architecture could also be applied to studying social dynamics, particularly those that involve the study of a biological substrate. In general, applying an LCS-inspired approach to evolutionary systems may open up new avenues of exploration, particularly in the study of evolutionary innovation, robustness, and cultural systems (see Supplemental Materials - Methods - Hybrid Fluid Dynamics-Evolutionary Agent model for latest information).

# Supplemental Information

## Methods

### Hybrid Fluid Dynamics-Evolutionary Agent Model

More information on details of the LCS-like model and how it maps to specific problems can be found in the following section. Included are technical details on flow field generation, genomic representations, replicator dynamics, and frontier analysis.

**Flow field.** The 2-dimensional flow field component of the CGS model is a metric space. Therefore, diffusion across this space can be measured and divergence between all particles can be directly compared. Each automata starts off in an initial cluster (shown in Figures A and B) and diffuses within the space according to the defined flow (Figure C).

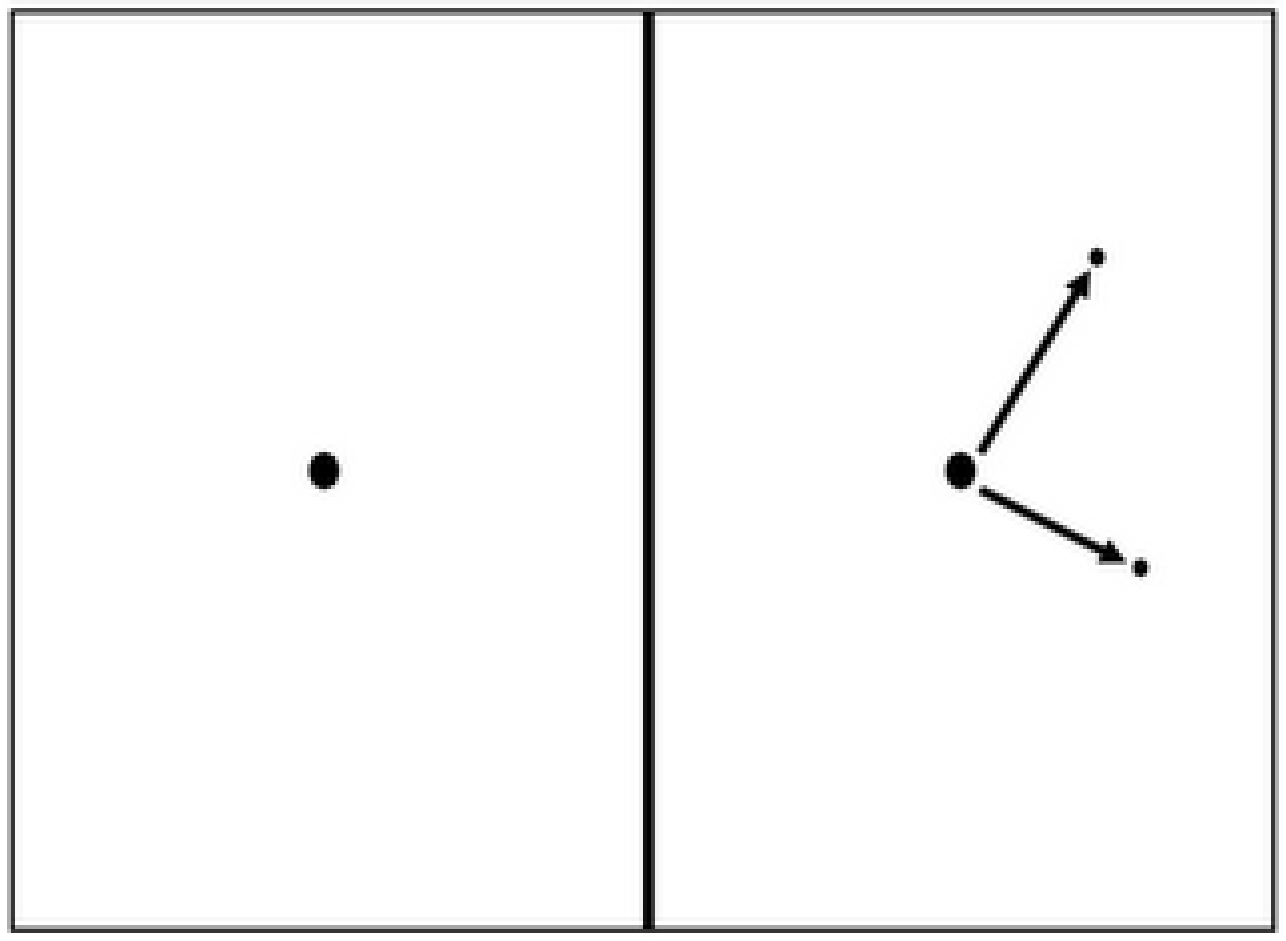

**Figure A. The divergence of two automata from a single initial population (cluster). In this case (laminar flow), the divergence is characterized as the arc between the two arrowheads. In a more turbulent flow, the paths of divergence would be less straight.**

The LCS-inspired flow field is generated using a set of flow rules. A flow field is required for diffusion of particles, and is a generalized stand-in for environment. In fluid dynamics, turbulent flows (or flows with large Reynolds number) produce the most interesting dynamics. In the model presented here, flow conditions can be created by using rulesets that specify the location, number, and strength of flows into the field.

Flow jets (shown in Figure C) are located along the field boundary, and the subsequent flows mediate the diffusion of particles. These flow conditions supervise the evolution of particles in the simulation, but depending on the turbulence produced do not lead to deterministic behavior. Figure D theoretically demonstrates the expected relationship between the degree of turbulence produced using a particular set of rules governing the location and strength of the flow jets. This graph also takes into account the number of step sizes taken by automata embedded in the flow field.

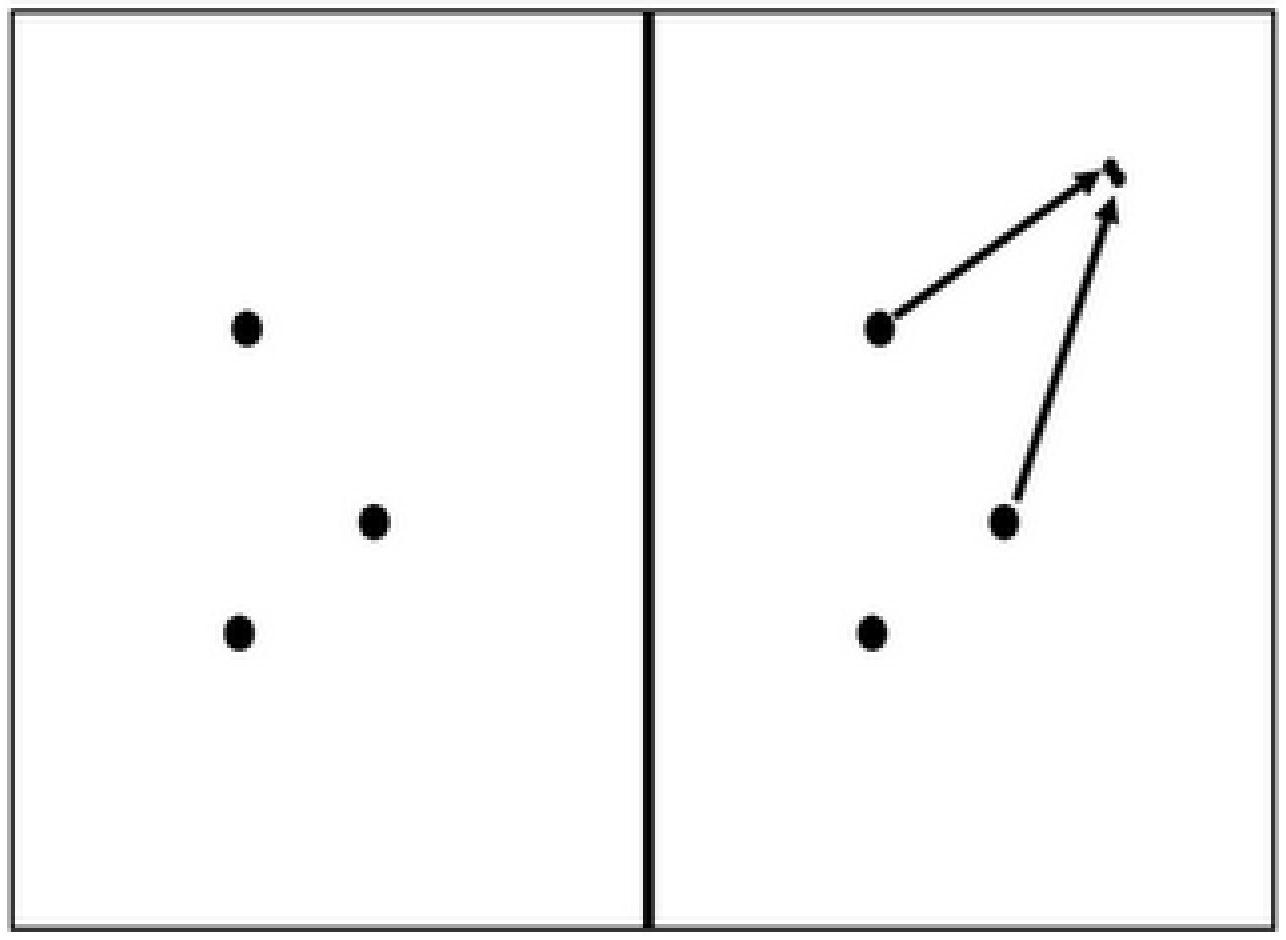

**Figure B. The convergence of two automata from different initial populations (clusters). The divengence is characterized as the arc between the two arrowheads. In this case, two automata starting out in different points have converged to a nearly identical position (negative divergence). This example also shows dynamic behavior under laminar flow conditions.**

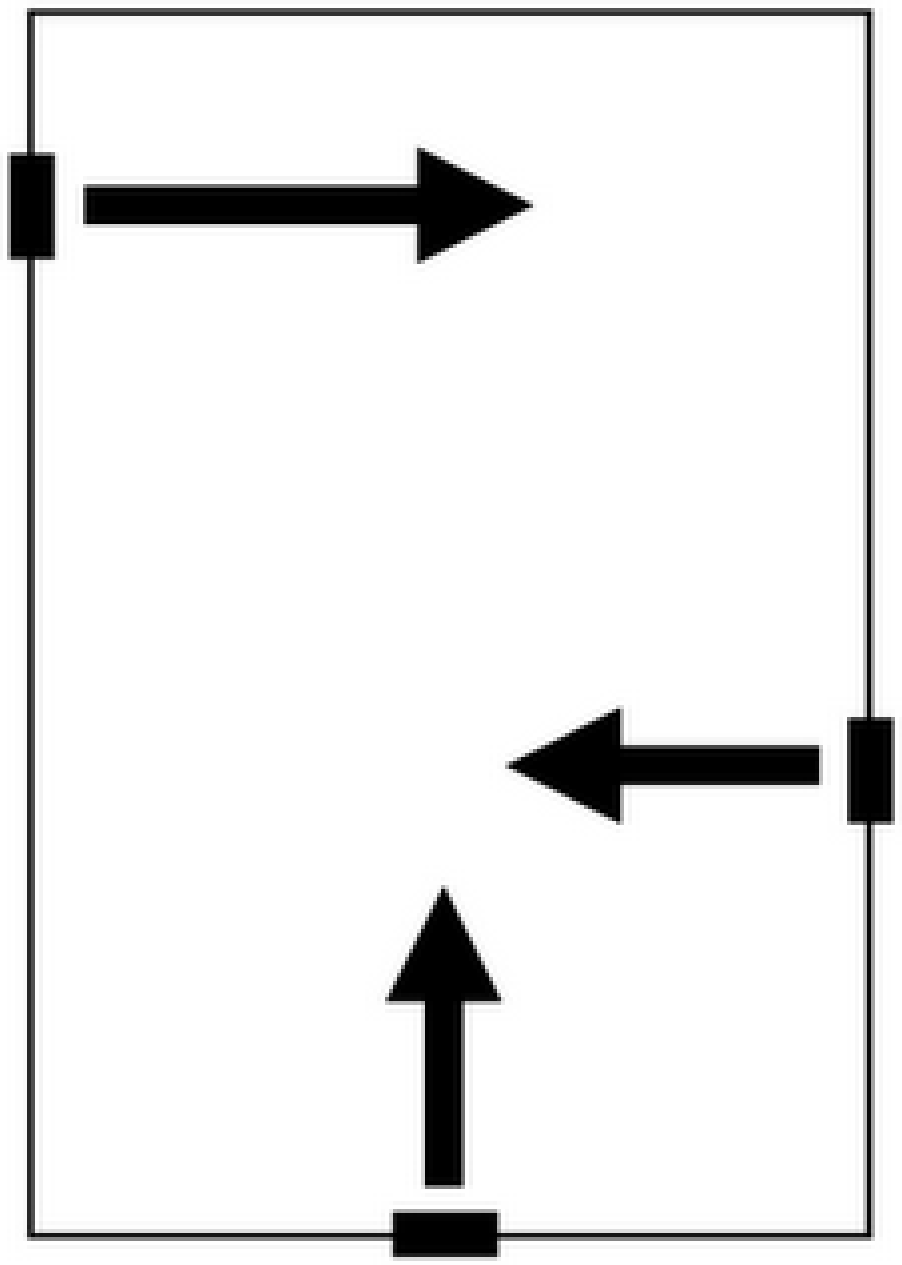

**Figure C. Flow jets (bars along boundary) and flows (arrows) for a 2-dimensional flow field.**

As we can see from the graph in Figure D, configurations that result in laminar flows (left side of graph) result in a consistent step size across the simulation. Meanwhile, configurations that result in highly turbulent flows (right side of graph) result in a highly variable step size both for the trajectory of single automata and the paths between automata. Since this is a physically-inspired model, this can be characterized by a degree of turbulence, which is proportional but not exactly equivalent to changes in Reynolds number [33].

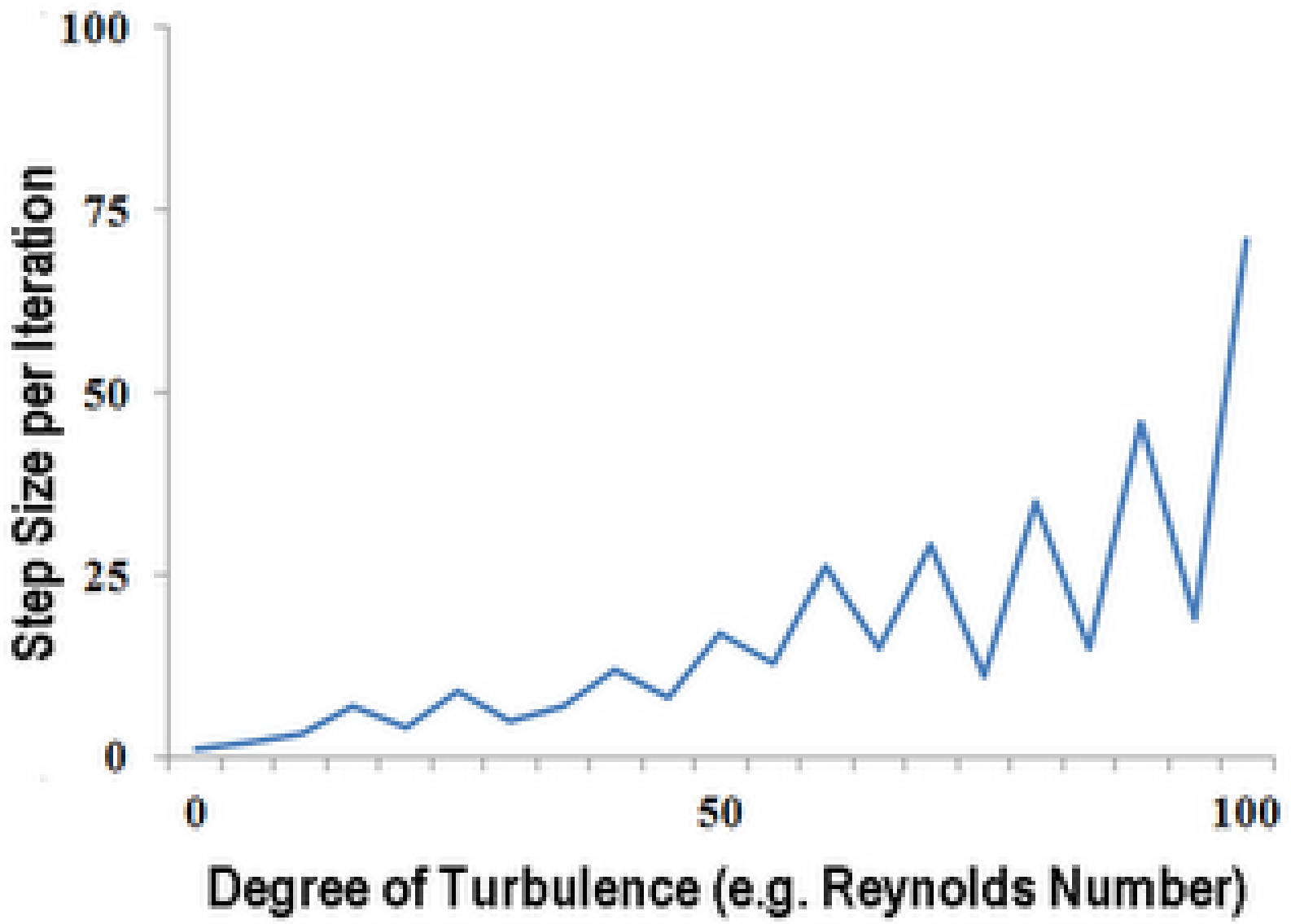

**Figure D. Graph (using pseudo-data) showing the relationship between the degree of turbulence created using the flow jets and the step size per iteration for automata embedded in the flow field.**

**Genomic representation.** All particles in a simulation contain a simple genomic representation similar to a chromosomal representation in a genetic algorithm. This genome exists primarily to identify the particle, and so consists of a short binary string that resembles a short tag. Each particle is initialized with a genome, and a single evolutionary epoch can commence with either a genetically homogeneous or genetically heterogeneous particle population. When a particle reproduces, it does so asexually, and passes its genome on to its offspring. Mutation is allowed, with a rate specified prior to initialization. The other function of a genome is to determine its response to environmental conditions (see Methods – Replicator Dynamics).

**Replicator Dynamics.** At randomly determined time intervals, perturbations are introduced that results in three potential outcomes for a given particle: mutate, die, or replicate (resulting in two children particles). The genome (see Methods – Genomic Representation) has a short series of elements (randomly generated) that determines the particle's resilience to environmental forces. If the environmental forces are above a predetermined threshold with respect to the survival measurement, the particle dies. If it is well below a lower-bound threshold, the particle will replicate without mutation. In all other cases, random mutation of the genome occurs. In a laminar flow, most if not all particles will replicate, while in a highly turbulent flow, particles will die.

*Survival measurement*. The survival measurement is a fitness-like measurement that accounts for the interaction between the flow field (environment - $\nabla E$) and the state of a particle's genome (G). Measurement of survival takes into account the local environmental turbulence for a given particle at time $t$, and compares it with the short series of elements that defines its survival capacity.

**Frontier Analysis.** A frontier analytical approach is appropriated to determine the boundaries of evolvability. In typical econometric frontier analysis, the goal is bivariate optimization. When plotted on a bivariate graph, the optimal points resulting from this exercise form a "frontier". In the LCS-like model, particle diffusion over time produces an optimal set of points along *n*-dimensions.

**Measures**

All measures are defined in this section. One of them (Iterated Temporal Divergence) is related specifically to the LCS-like model, while the other two (Segregation Factor, and Conditional Diversity) are related to evolutionary dynamics themselves. The latter two measurements are adapted from Alicea (2012).

**Iterated Temporal Divergence (ITD).** Iterated Temporal Divergence is defined using the following equation

$$L_t(X_0) = \int_t^{t+1} (\nabla - v) | F_t^s (X_0) ds \quad [1]$$

where the divergence between two particles subject to the same flow field is integrated over a finite time period, t: $\rightarrow$ t + 1.

**Segregation Factor.** The segregation factor is used to understand changes in the distribution of values for a particular particle genome. Particles populations can become segregated over time, resulting from interactions with other particles and the flow field itself. This can be defined as

$$S = \left|\sum I_{ij\ldots n}\right| \ , \ \left|\sum I_{ij\ldots n}\right| > 0 \quad [2]$$

where a value of S $\rightarrow$ $S_{max}$ results in a maximization of movement towards discrete positions in the *n*-dimensional volume.

**Conditional Diversity.** To measure the distribution of automata within a given ridge or vortex, we can use a measure of conditional diversity. This measure provides us with a distribution of automata in the flow field for all automata within a certain value of the ITD measure (see equ. [1]). This measure can be stated as

$$D = \sigma\,(p_1, p_2, \ldots\ldots, p_n)$$

$$p_i = \frac{A_i}{A_{tot}} \quad [3]$$

$$A_i = \underset{X_0}{argmax}\ L_t X_0 \leq L_t X_0 \geq 0$$

where $\sigma$ equals the variance of set $p_n$, $A_i$ equals all automata for a specific subpopulation below the threshold value for the ITD measure, $p_i$ is the number of automata in a specific subpopulation, $A_{tot}$ is the total number of automata, and $p_n$ is the number of subpopulations in the simulation.

## SUPPLEMENTAL FIGURES

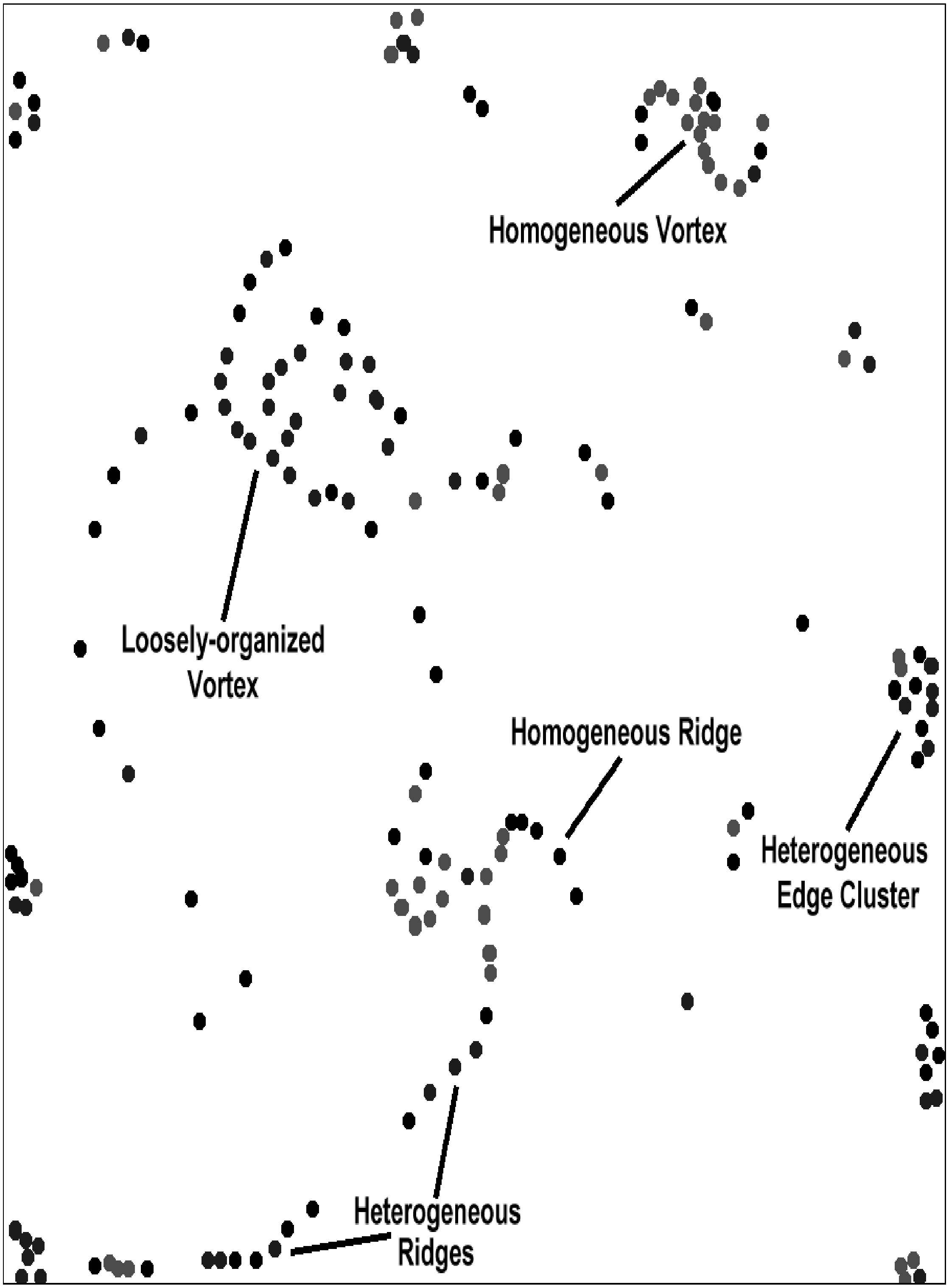

**Supplemental Figure 1. A 2-dimensional space representing an evolved population of automata representing distinct genotypes (gray and black). Each subpopulation has a multifaceted set of relationships with regard to the other. Taken from Figure 4, [18].**

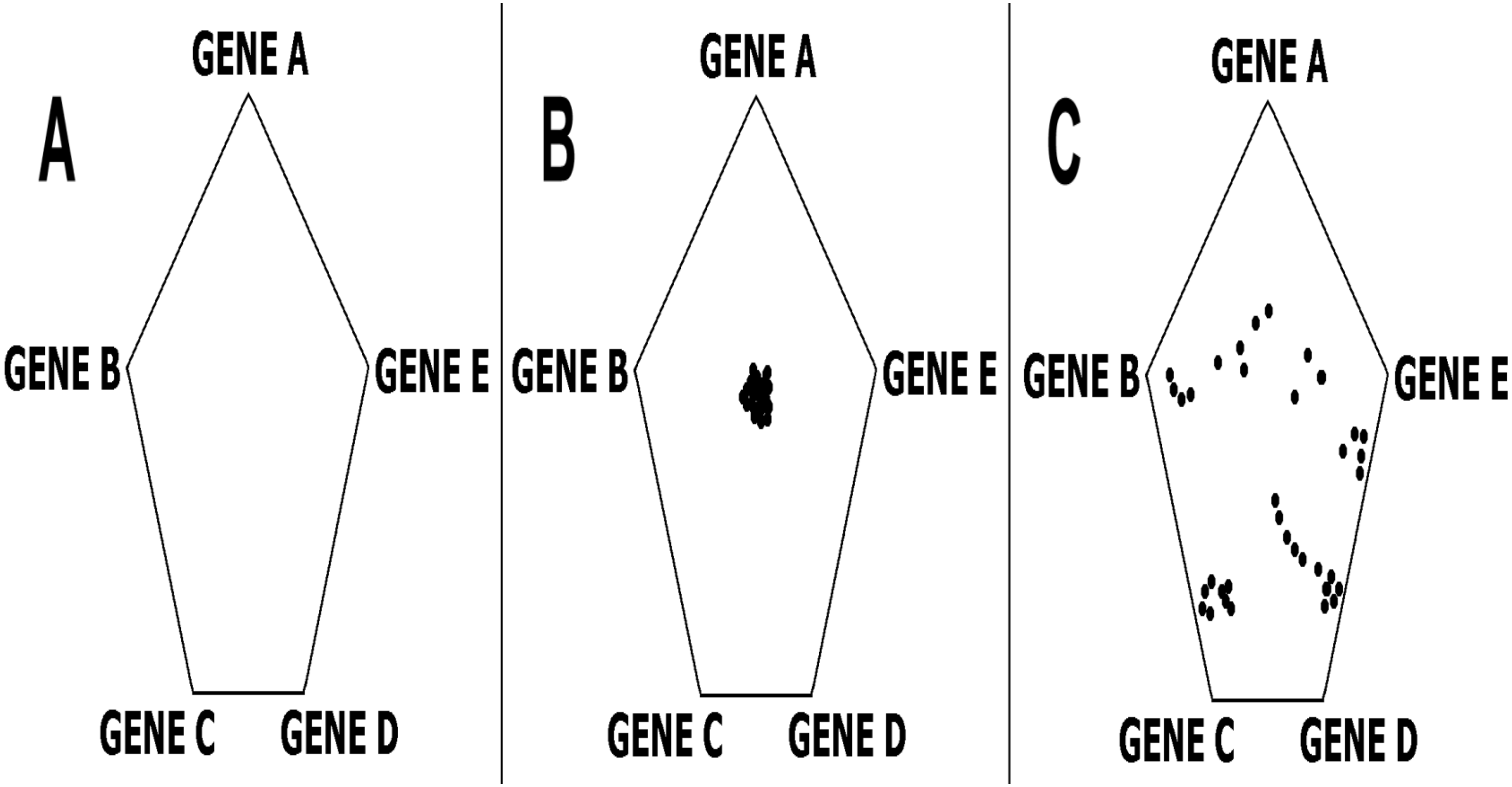

**Supplemental Figure 2. Organismal trait metric space based on volume shown in Figure 5. A: volume before particle seeding, B: volume initial condition, C: volume after a single evolutionary epoch (*n* generations).**

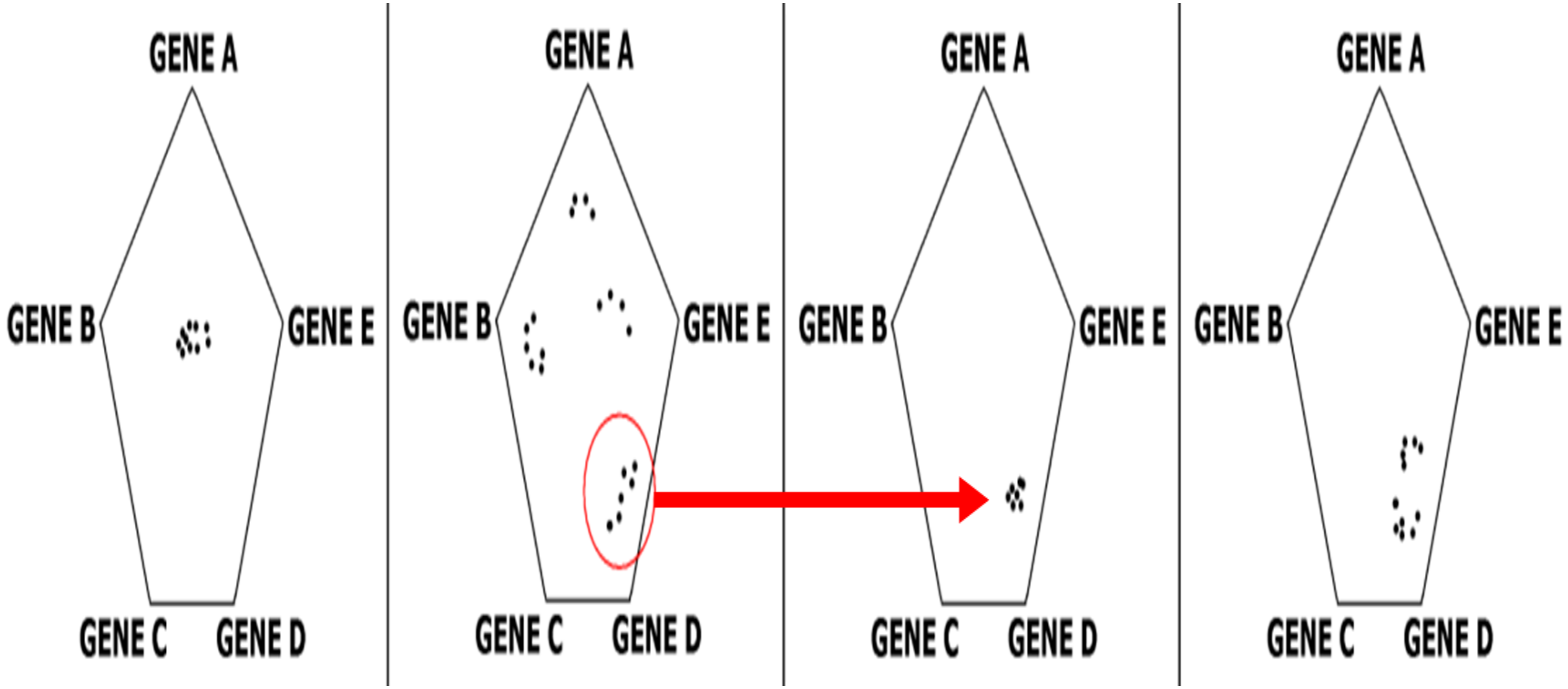

**Supplemental Figure 3. An example of an LCS-like model used to approximate a demographic bottleneck. FROM LEFT: a 5-dimensional space is seeded with agents, which are allowed to evolve (diffuse). A subpopulation of agents are then resampled, and used to seed an empty 5-dimensional space. The process can be infinitely recursive for a given set of genotypes.**

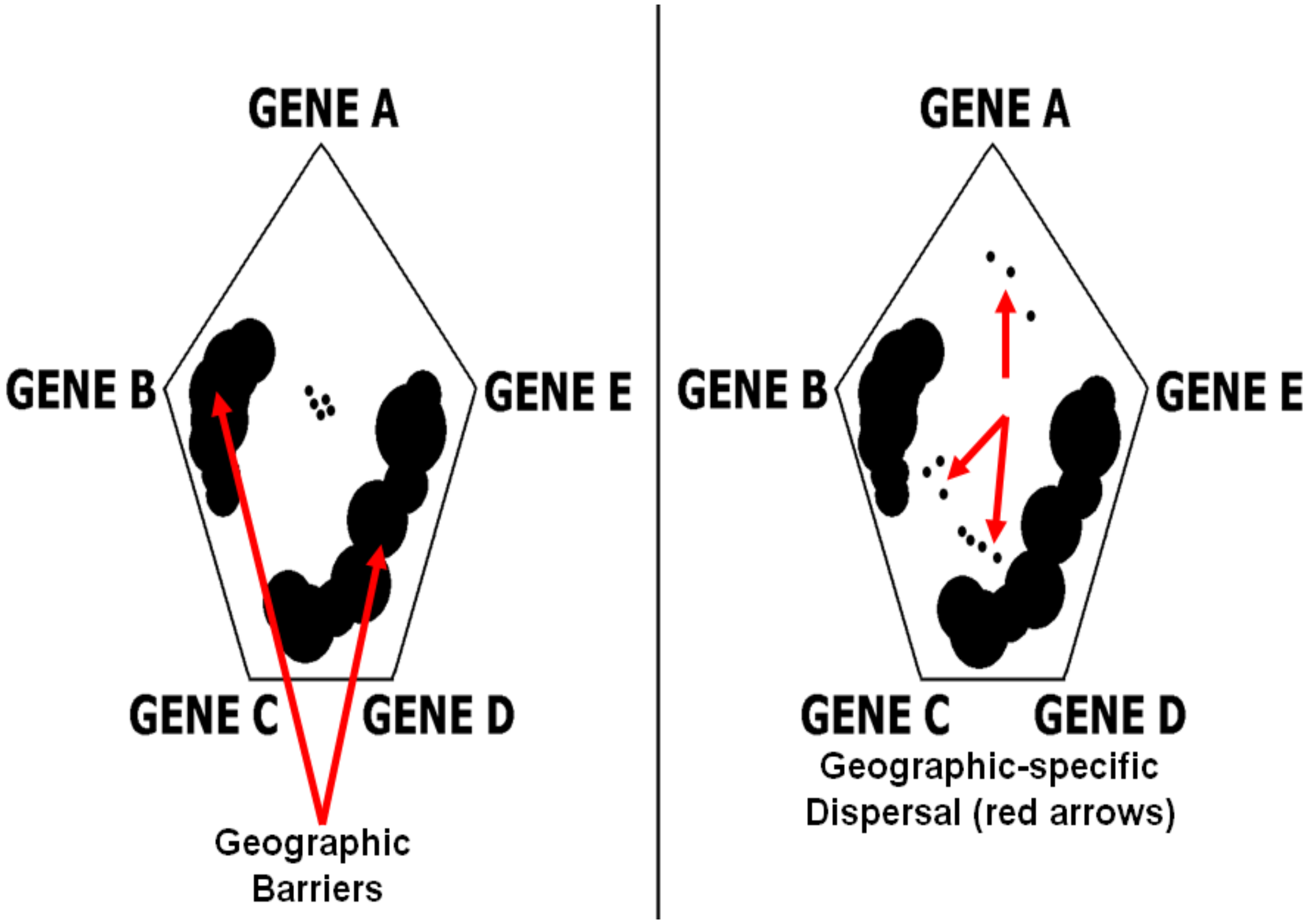

**Supplemental Figure 4. An example of an LCS-like model used to approximate biogeography. In this example, a 5-dimensional space is seeded with agents and geographic barriers. The agents are then allowed to diffuse, but are constrained by the static geographic barriers.**